\documentclass[twocolumn,twocolappendix,trackchanges]{aastex63}
\usepackage{bm}
\usepackage{braket}
\usepackage{amsmath}
\usepackage{subeqnarray}
\usepackage{here}

\received{}
\revised{}
\accepted{}
\submitjournal{ApJ}
\shorttitle{Angular momentum transport in the convection zone}
\shortauthors{Shimada et al.}
\begin{document}
\title{Angular momentum transport in the convection zone of a 3D MHD simulation of a rapidly rotating core-collapse progenitor}

\correspondingauthor{Ryota Shimada}
\email{shimada@kusastro.kyoto-u.ac.jp}

\author[0000-0002-8507-3633]{Ryota Shimada}
\affiliation{\rm{Department of Astronomy, Kyoto University, Sakyo-ku, Kyoto, 606-8502, Japan}}

\author[0000-0003-1752-4882]{Lucy O. McNeill}
\affiliation{\rm{Department of Astronomy, Kyoto University, Sakyo-ku, Kyoto, 606-8502, Japan}}
\affiliation{\rm{Hakubi Centre for Advanced Research, Kyoto University, Kyoto 606-8317, Japan}}\
\affiliation{\rm{Center for Interdisciplinary Theoretical and Mathematical Sciences (iTHEMS), RIKEN, Saitama 351-0198, Japan}}

\author[0000-0002-1685-2465]{Vishnu Varma}
\affiliation{\rm{Astrophysics Group, Lennard-Jones Laboratories, Keele University, Keele ST5 5BG, UK}}

\author[0000-0003-2611-7269]{Keiichi Maeda}
\affiliation{\rm{Department of Astronomy, Kyoto University, Sakyo-ku, Kyoto, 606-8502, Japan}}

\author[0000-0001-5457-4999]{Takaaki Yokoyama}
\affiliation{\rm{Astronomical Observatory, Kyoto University, Sakyo-ku, Kyoto, 606-8502, Japan}}

\author[0000-0002-4470-1277]{Bernhard M\"uller}
\affiliation{\rm{School of Physics and Astronomy, 10 College Walk, Monash University, Clayton, VIC 3800, Australia}}





\begin{abstract}
Rotation and magnetic fields in the cores of evolved massive stars in their final phase are thought to play an important role in the subsequent supernova explosion and the formation of a compact object, especially in hyperenergetic explosions. However, the interplay between rotation, magnetic fields, and convection up to the final collapse is a nonlinear, multidimensional effect that is difficult to capture with standard one-dimensional (1D) stellar evolution models. We quantify the magnetic angular momentum (AM) transport in the convective oxygen burning shell in a three-dimensional (3D) rotating core-collapse progenitor model.
We find that the radial direction of magnetic AM transport is directly related to the Rossby number of the convective oxygen shell.
We also analyze the magnetic energy, which sets the amplitude of the magnetic AM flux.
The magnetic energy is determined both by rotation and the nuclear energy generation rate analogously to low-mass stars like the Sun.
Based on these results, we construct a 1D model of magnetic AM transport in the convection zone for the first time in terms of properties of a given stellar evolution model.
This model successfully reproduces the AM transport in the 3D model when the convective dynamo is in a quasi-steady state.
Notably, our model for radial AM transport is the first to account for inward AM flux. 
This may result in interesting differences compared to the conventional treatment of magnetic AM transport in stellar evolution models, which assume AM is transported outward by a purely diffusive process. 
\end{abstract}


\section{Introduction\label{sec:intro}} 
Rotation and magnetic fields in the cores of massive star immediately prior to core-collapse are key factors determining the nature of the explosion and the properties of the resulting compact object \citep{Wheeler_2002, Heger_2005}.
There is now an increasing consensus that efficient angular momentum transport usually results in relatively slow pre-collapse rotation \citep{Heger_2005,Fuller_2019} in line with the observed birth spin periods of pulsars. However, hyperenergetic core-collapse supernovae of Type Ic-BL (broad-lined) are difficult to account for without rapidly rotating stellar cores
as a prerequisite for magnetorotational explosions
\citep[e.g.,][]{akiyama_03,Mosta_2014,Obergaulinger_2021,mueller_25}.
Evolutionary pathways for the formation of hyerpnova progenitors
with rapidly rotating cores have been proposed \citep{Woosley_06,aguilera_18},
but these rely on recipes for angular momentum (AM) transport and field amplification within spherically symmetric stellar evolution models. Multi-dimensional magnetohydrodynamic (MHD) simulations of stellar interiors over evolutionary time scales
could more accurately capture the nonlinear interaction between the rotation, magnetic fields, and convection, but remain computationally unfeasible. For this reason, the evolution of the rotation and magnetic fields towards core-collapse remains highly uncertain. However, 3D MHD simulations of stellar interiors on short time scales are starting to shed light on the intricate problem of AM transport and magnetic field amplification.
\\

Recent 3D core-collapse progenitor simulations by \citet{Varma_2023} have demonstrated that strong magnetic fields, which are generated in convective shells during the late stages of stellar evolution, can efficiently transport AM from the O shell burning region surrounding the core to the outer layers before collapse.
These results demonstrate the importance of magnetic AM transport in determining the structure of the progenitor stars and underscore shortcomings of current stellar evolution models.
\\

Effective models for AM transport in 1D stellar evolution models typically treat hydrodynamic (HD) effects of rotationally induced circulations, instabilities \citep[e.g.][]{Meynet_2000}.
These models were later extended to incorporate magnetic torques in radiative zones.
\citet{Spruit_2002} proposed that dynamo field amplification
based on the Tayler instability \citep{tayler_73}
predominantly generates horizontal magnetic fields, enabling more efficient angular momentum transport than purely hydrodynamic processes \citep[][]{Spruit_2002,Fuller_2019}. 
These prescriptions have been implemented into 1D stellar evolution codes \citep[e.g.,][]{Heger_2005,Eggenberger_2008,Paxton_2013} to model the evolution of core rotation from the main sequence through core collapse.
\\

Observationally, rotation rates of stellar cores are measured with asteroseismology, providing observational constraints on 1D AM transport models \citep{Aerts_2019}.
Asteroseismic observations reveal significant discrepancies between theoretical predictions and stellar rotation rates across a diversity of stellar populations.
For example, Kepler observations of 1–2\,$M_\odot$ red giants reveal core rotation periods of 10–100 days \citep[e.g.][]{Cantiello_2014}, which are much longer than the $\sim$1 day periods predicted by purely hydrodynamic models \citep[e.g.][]{Moyano_2022}.
In other words, purely 1D hydrodynamic (HD) models overestimate core rotation rates by nearly two orders of magnitude \citep{Moyano_2022}.
Similarly, observations of pulsating white dwarfs indicate rotation rates far below those predicted by 1D models \citep{Hartogh_2019}.
\\

The various discrepancies between asteroseismically measured rotation rates and predictions from stellar evolution models suggest that a key mechanism for AM transport is missing from standard stellar evolution prescriptions.
One potential means to extract AM from the core is via magnetic torques.
For example, \citet{Fuller_2019} suggested a modified Tayler–Spruit dynamo model for radiative zones, which can reproduce the core rotation rates of red giants and white dwarfs under certain assumptions.
However, to date, there is no evolutionary model that adequately models the convective dynamo across various stellar populations, or over dynamical timescales.
\\

The solar convection zone constitutes one example where 3D magnetoconvection simulations have been extensively performed and can also be compared with detailed observational data. In particular, 3D simulations of solar-like stars demonstrate how small-scale turbulent magneto-convection can influence the rotation rate, i.e., they affect AM transport by small-scale turbulent correlations, namely Reynolds and Maxwell stresses.
For example, the rotationally-induced anisotropy in turbulence has been shown to generate Reynolds stresses that are crucial for maintaining large-scale differential rotation profiles in the solar convection zone \citep{Kitchatinov_1993, Miesch_2005}.
Recent high-resolution 3D MHD simulations further reveal that small-scale magnetic field correlations, i.e., Maxwell stresses, dominate radial AM transport, enabling the formation of the solar-like differential rotation, i.e., the equator rotating faster \citep{Hotta_2022}.
Moreover, \citet{Augustson_2019} demonstrated, through both semi-analytic theory and 3D MHD simulations, that the ratio of magnetic to kinetic energy (ME/KE) increases with the increasing dynamical influence of rotation over inertia.
This indicates a stronger magnetic feedback in the low Rossby number case of rapidly rotating convection.
In extreme cases, ME/KE can exceed unity at low Rossby numbers, so that magnetic AM transport becomes crucial.
\\

For massive stars during advanced burning phases, recent 3D HD simulations \citep{McNeill_2022} reveal radial and latitudinal angular-velocity gradients within the oxygen-burning shell. 
Interestingly, rotation in the convective regions is reminiscent of solar differential rotation.
In addition, they find that AM transported by convection is not well described by the standard model by the 1D mixing length theory (MLT) \citep{Bohm_1958}.
Specifically, MLT fails to predict both the magnitude and direction of convective AM transport.
Nevertheless, the role of magnetic fields in AM transport within convective zones of massive stars remains largely unexplored.
\\

The aforementioned 3D MHD simulation by \citet{Varma_2023} demonstrates the magnetic amplification and AM transport by convective dynamo action in the cores of massive stars. Their simulation of convective shells in a rotating $16\,M_\odot$ progenitor during the oxygen and neon-burning phases show that convective dynamo action amplifies magnetic fields to strengths of $\sim10^{11}$\,G. 
These strong fields generate Maxwell stresses that are comparable to, or even exceed, the Reynolds stresses. 
As a result, convection is suppressed, while AM is transported efficiently outward. 
Consequently, the burning shells rapidly evolve toward near-rigid rotation. 
Such magnetic braking by the convective dynamo, which is absent in standard 1D stellar evolution models, suggests that the 1D models may substantially overestimate core rotation rates at collapse. 
\\

Although the Maxwell stress in radiative zones can in principle be described by 1D models, no comparable formulation has yet been established for convective zones.  
In this work, we aim to develop a 1D model for the Maxwell stress applicable to convective regions in stellar evolution.
To do this, we analyze the aforementioned 3D MHD simulation by \citet{Varma_2023}.
Section~\ref{sec:Motivation} reviews the physics of AM transport in convective zones and the 1D approximation for radial AM transport used in stellar evolution models.
Section~\ref{sec:results} summarizes the simulation setup and presents an analysis of physics which determines both the direction and the amplitudes of the magnetic AM transport.
In Section~\ref{sec:model}, we construct the 1D magnetic AM transport model based on our analysis and compare it with the simulation. 
Section~\ref{sec:Discussions} discusses the implications and limitations of our model and suggest future analyzes.
Section~\ref{sec:conclusion} summarizes our conclusions.
   
\section{ANGULAR MOMENTUM TRANSPORT IN STELLAR CONVECTION ZONES}\label{sec:Motivation}
To develop a physically motivated prescription for magnetic AM transport applicable to stellar evolution models, it is essential to clarify the underlying physics of AM redistribution within convective zones. In this section, we review the theoretical and numerical foundations of AM transport in stellar convection and evolution.  
We begin by reviewing the mechanisms of AM transport driven by rotating magnetoconvection, as revealed in simulations of solar-type stars (Section~\ref{subsec:Rossby_AM}).  
Next, in Section~\ref{subsec:AM_equation} we introduce the governing equations for radial AM transport used in 1D and 3D stellar evolution simulations. Then in Section~\ref{subsec:AM_evo} we summarize the physically motivated approximations for convective AM transport commonly used in 1D stellar evolution models.  
Finally in Section~\ref{subsec:AM_mag} we review the 1D formulation of magnetic AM transport in radiative zones.

\subsection{Rossby number dependence of turbulent angular momentum transport}\label{subsec:Rossby_AM}
 Internal AM transport has long been discussed in the context of the solar convection zone \citep[e.g.,][]{Busse_1970, Gilman_1979, Kitchatinov_1993, Miesch_2000}, 
where the rotation profile in the meridional plane is well constrained by helioseismic observations \citep[e.g.,][]{Howe_2009}.  
In particular, both 3D HD and MHD simulations have demonstrated that turbulent AM transport strongly depends on the convective Rossby number $\mathrm{Ro}$ \citep[e.g.,][]{Miesch_2005}, defined as
\begin{eqnarray}
\displaystyle &&\mathrm{Ro}=\frac{v_{\mathrm{conv}}}{2\Omega L}, \label{eq:Def_Ro}
\end{eqnarray}
where $\Omega$, $v_{\mathrm{conv}}$, and $L$ denote the rotation rate, convective velocity, and characteristic length scale, respectively.  
A small Rossby number indicates a regime where the Coriolis force dominates over advection.
In such cases, the Coriolis force acting on radial flows creates a positive correlation between radial and longitudinal velocities ($v_r$ and $v_\phi$, respectively), concentrating AM to equatorial regions.
This correlation ultimately transports AM outward via the Reynolds stress.
\\

In addition, recent high-resolution 3D MHD simulations of the solar convection zone \citep{Hotta_2022} suggest that AM transport by small-scale magnetic fields plays a crucial role in maintaining solar differential rotation, which is faster at the equator despite a large Ro.  
Specifically, these simulations also reveal a close relationship between turbulent (Reynolds stress) and magnetic (Maxwell stress) AM transport through an alignment between plasma flow and magnetic fields.  
In the framework of ideal MHD, magnetic field lines are effectively “frozen” into the moving plasma \citep{Alfven_1943}.  
This frozen-in condition causes the magnetic field lines to remain tightly coupled to the plasma flow, and the magnetic fields vector $\bm{B}$ is aligned with fluid velocity vector $\bm{v}$ nearly parallel or antiparallel.
As a result, the Reynolds and Maxwell stresses are strongly interdependent. 3D solar convection zone simulations therefore suggest that the Rossby number also plays a key role in determining the efficiency of magnetic AM transport.
\\

Moreover, both observations and theoretical studies have shown that the efficiency of convective dynamos strongly depends on the Rossby number in low-mass stars \citep{See_2019, Reiners_2022, Augustson_2019, Brun_2022}.
Specifically, this Rossby number dependence predicts stronger magnetic fields at lower Rossby numbers.
Consequently, enhanced magnetic fields produces significant AM flux by the Maxwell stress due to its interdependence with the Reynolds stress.

\subsection{Angular momentum transport equation}\label{subsec:AM_equation}
The evolution equation for AM can be obtained by taking the cross product of the position vector with the momentum equation of the MHD system.  
If one is interested in the radial transport of AM, it is sufficient to consider the radial AM flux obtained through a spherical average.  
This flux can be expressed as
\begin{eqnarray}
  \displaystyle 
  \frac{\partial \braket{\rho v_\phi r \sin\theta}}{\partial t}
  +\nabla_r\cdot\Braket{\rho v_r v_\phi r \sin \theta - \frac{B_rB_\phi}{4\pi}r\sin\theta}
  =0, \nonumber\\
  \
\end{eqnarray}
where $\braket{}$ and $\nabla_r\cdot$ denote the spherical averaging operation and the radial component of the divergence operator, respectively.  
\\

In systems where the background stratification evolves dynamically, the Favre decomposition \citep{Favre_1965} provides a useful method for isolating the physical mechanisms of turbulent convection.  
This technique has also been applied to analyses of AM transport in 3D HD simulations of massive stars during oxygen (O) shell burning \citep{McNeill_2020,McNeill_2022,Fields_2022}.  
Under a Favre decomposition, the evolution equation for AM in an MHD system can be written as
\begin{eqnarray}
  \displaystyle
  \frac{\partial \hat{\rho}\tilde{\Omega}_z\tilde{i}_{zz}}{\partial t}
  &&+\nabla_r\cdot \Big(
    \braket{\rho\tilde{v}_r(\tilde{\Omega}_z+\Omega''_z)r^2\sin^2\theta}
    +\braket{\rho v''_r\tilde{\Omega}_z r^2\sin\theta} \nonumber\\
  &&+\braket{\rho v''_r\Omega''_z r^2\sin^2\theta}
    -\Braket{\frac{B_rB_\phi}{4\pi} r\sin\theta}
  \Big)=0, \label{eq:AM_Con}
\end{eqnarray}
where $\tilde{\Omega}_z$ and $\tilde{i}_{zz}$ are the Favre averaged rotation rate and the $zz$-component of the gyration tensor, respectively.  
Details of the Favre decomposition are given in Appendix \ref{sec:appendix A}.  
\\

The second term in Eq.~(\ref{eq:AM_Con}) represents AM transport due to advection by the large-scale flow ($\tilde{v}_r$), i.e., contraction or expansion.  
The third term corresponds to meridional circulation, the fourth term accounts for turbulent transport (Reynolds stress), and the fifth term describes magnetic transport (Maxwell stress). 

\subsection{Angular momentum transport in stellar evolution models}\label{subsec:AM_evo}
 
On the actual stellar evolutionary timescale, it is computationally expensive to compute the angular momentum evolution directly with 3D simulations.
Therefore, Eq.~(\ref{eq:AM_Con}) is simplified by the models that describe only the radial transport (1D approach).
In such models, meridional circulation, Reynolds stress, and Maxwell stress cannot be solved from first principles, so that some simplifying assumptions (closures) must be introduced.  
\\

As a representative example, \citet{Maeder_1998} proposed the following 1D HD model for AM transport:
\begin{eqnarray}
  \displaystyle 
 &&\frac{\partial \rho r^2\Omega_z}{\partial t}
 +\frac{1}{r^2}\frac{\partial }{\partial r}(\rho v_r\Omega_z r^4)
 -\frac{1}{r^2}\frac{\partial}{\partial r}\left(\rho r^4\frac{1}{5}\Omega_zv_Q\right)\nonumber\\
 &&-\frac{1}{r^2}\frac{\partial}{\partial r}\left(\rho r^4D_\mathrm{conv}\frac{\partial \Omega_z}{\partial r}\right)=0,\label{eq:1D_Maeder}
\end{eqnarray}
where $\Omega_z$ is rotation rate. Each term in Eq.~(\ref{eq:1D_Maeder}) approximates the respective term in Eq.~(\ref{eq:AM_Con}) excluding the final Maxwell stress term. In the third term, which characterizes the polar to equator transport of AM, $v_Q$ is the velocity of the quadrupolar meridional circulation. In the fourth term, which describes the turbulent transport of AM via Reynolds stress, $D_{\mathrm{conv}}$ is the convective diffusion coefficient. This is typically modeled in the framework of mixing length theory (MLT) \citep[][]{Bohm_1958}.
\\ 

In the framework of MLT, convective transport as a diffusive process, along with the corresponding diffusion coefficient, can be expressed in terms of the convective velocity ($v_\mathrm{conv}$) and the pressure scale height ($H_p$) as follows:
\begin{eqnarray}
  \displaystyle 
  &&D_\mathrm{conv}=\frac{1}{3}v_\mathrm{conv}H_p.
\end{eqnarray}
The convective velocity can be estimated from the degree of the convective instability of the stratification \citep[][]{Kippenhahn_1990}.  
One of the common measures for the criticality of convective instability is superadiabaticity \citep[][]{Kippenhahn_1990}, which we refer to as SA, defined as:
\begin{eqnarray}
\displaystyle \mathrm{SA} &&\equiv \left(\frac{\partial \ln T}{\partial \ln p}\right)
-\left(\frac{\partial \ln T}{\partial \ln p}\right)_{\mathrm{ad}}
-\frac{\varphi}{\delta}\left(\frac{\partial \ln \mu}{\partial \ln p}\right)\nonumber\\
&&=-\frac{H_p}{C_p}\frac{dS}{dr}
-\frac{\varphi}{\delta}\left(\frac{\partial \ln \mu}{\partial \ln p}\right) , \label{eq:SA}
\end{eqnarray}
where $\varphi$, $\delta$, and $C_p$ are $(\partial \ln \rho/\partial \ln \mu)_{p,T}$, $-(\partial \ln \rho/\partial \ln T)_{p,\mu}$, and the specific heat at constant pressure, respectively.
$S$ and $\mu$ denote the specific entropy and the mean molecular weight. 
A positive $\mathrm{SA}$ indicates that the region is convectively unstable.
Note that there are equivalent criteria to Eq.~(\ref{eq:SA}) for convective instability based on, e.g., the Brunt-V\"ais\"al\"a frequency or the entropy gradient.
They are physically equivalent descriptions of the same underlying process.
In MLT, the balance between buoyant acceleration and inertia yields the convective velocity in terms of the superadiabaticity as:
\begin{eqnarray}
  \displaystyle 
  &&v_\mathrm{conv}^2=g\delta\frac{\alpha_{\mathrm{MLT}}^2H_p}{8}\mathrm{SA}.\label{eq:v_MLT}
\end{eqnarray}
$\alpha_{\mathrm{MLT}}$ is the mixing-length parameter, typically of order unity. 
$g$ is the gravitational acceleration. While the diffusion coefficient $D_\mathrm{conv}$ in Eq.~(\ref{eq:1D_Maeder}) describes AM transport by convection, \citet{Heger_2000} generalized diffusive AM transport to cover a wider range of hydrodynamic instabilities. This includes diffusive AM transport from the effects of secular shear, dynamical shear, Eddington–Sweet circulation, and the Goldreich–Schubert–Fricke instability. This comprehensive approach allowed their model to treat AM transport even outside the convection zone.

\subsection{Angular momentum transport by magnetic fields}\label{subsec:AM_mag}
 
The modeling of the Maxwell stress term in Eq.~(\ref{eq:AM_Con}) has so far been limited to magnetic fields generated by the Tayler–Spruit dynamo in convectively stable radiative zones.  
Originally proposed by \citet{Spruit_2002}, this formulation was later extended by \citet{Fuller_2019} to include the nonlinear saturation process. In \citet{Fuller_2019}, the Maxwell stress is also represented as a diffusive process, analogous to the fourth term in Eq.~(\ref{eq:1D_Maeder}), and the corresponding diffusion coefficient is estimated as
\begin{eqnarray}
  \displaystyle 
  &&D_{\mathrm{TS}}=\alpha_{\mathrm{TS}}^3 r^2\Omega_z\left(\frac{\Omega_z}{N_{\mathrm{eff}}}\right)^2,
\end{eqnarray}
where $\alpha_{\mathrm{TS}}$ is a parameter of order unity and $N_{\mathrm{eff}}$ denotes the effective Brunt-V\"ais\"al\"a frequency.
Because the radiative zone in late evolutionary stages generally rotates faster in its inner regions, the resulting Maxwell stress therefore always transports AM outward. Although the Maxwell stresses in radiative zones can be described by such 1D models, it is still limited to convectively stable regions. No comparable formulation has yet been established for convective zones.  
In this work, we aim to develop a 1D model for the Maxwell stresses applicable to convective regions.
To do this, we will analyze a 3D MHD simulation.
\\

\section{RESULTS}\label{sec:results}

Building on the theoretical background developed in Section~\ref{sec:Motivation}, this section investigates the properties of magnetic AM flux within the convective zones of a core-collapse progenitor.  
We provide an overview of the 3D MHD simulation (Section~\ref{subsec:num_set}) and describe the evolution of the rotation rate obtained in the simulation (Section~\ref{subsec:Evolution of angular momentum in the simulation}).
Section~\ref{subsec:setup} defines the spatial and temporal domains for our analysis, and Sections \ref{subsec:dirct_MS} and \ref{subsec:results_amplitudes} provide the detailed analysis on magnetic AM transport.

\subsection{3D MHD simulation of a rapidly rotating core-collapse progenitor}\label{subsec:num_set}
We take the 3D MHD simulation of convective oxygen, neon, and carbon shell burning in a rapidly rotating $16\,M_\odot$ helium star presented by \citet{Varma_2023}. The progenitor model originates from the Kepler stellar evolution calculations of \citet{Woosley_06}. It features strong radial differential rotation in radiative zone and has also been used in the 3D HD shell convection study of \citet{McNeill_2022}.
In this low Rossby number regime, they obtained latitudinal differential rotation which rotates fastest at the equator, which is consistent with the relationship between Reynolds stress and Rossby number found in the low mass star context (Section~\ref{subsec:Rossby_AM}).
\\

The simulation of \citet{Varma_2023} was performed using the Newtonian MHD version of the CoCoNuT code \citep{Mueller_2020, Varma_2021} in spherical polar coordinates with full $4\pi$ coverage in latitude and longitude.  
The computational domain extends from an inner radius of 3,000\,km, where the core is replaced by a point mass, to an outer radius of 40,000\,km, encompassing the O, Ne, and C-burning shells.  
The initial magnetic field is homogeneous and aligned with the rotational axis and has a strength of $10^7$\,G.  
 After approximately 10 convective turnover times, this is subsequently amplified to $10^{10}$ – $10^{11}$\,G in O and Ne-burning shells. In this work, we concentrate on the angular-momentum evolution of the O-burning shell, which is the most spatially well-resolved region in our data owing to the logarithmic spacing of the radial grid.

\subsection{Evolution of angular momentum in the simulation}\label{subsec:Evolution of angular momentum in the simulation}
\begin{figure*}
 \includegraphics[width=\textwidth]{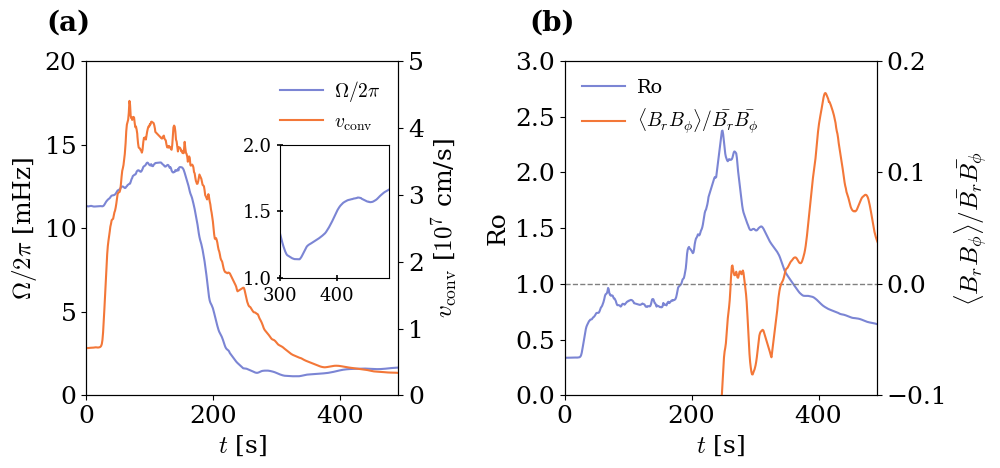}
 \caption{(a) Temporal evolution of Favre averaged rotation rate $\Omega/2\pi$ (blue solid line) and convective velocity $v_\mathrm{conv}$ (orange solid line) in the O shell. 
 The vertical axes for the rotation rate and convective velocity are displayed on the left and right sides, respectively.
 An inset view on the right side highlights the evolution of the rotation rate $\Omega/2\pi$ during the quasi-steady phase after $300$\,s, where the system maintains a statistically steady convective dynamo.
 (b) Temporal evolution of the Rossby number Ro (blue solid line) and the correlation of Maxwell stress (orange solid line) in the O shell.
 The vertical axes for the Rossby number and the correlation of Maxwell stress are displayed on the left and right sides, respectively.}
 \label{fig:Ro_def}
\end{figure*}

\begin{figure}
 \includegraphics[width=\columnwidth]{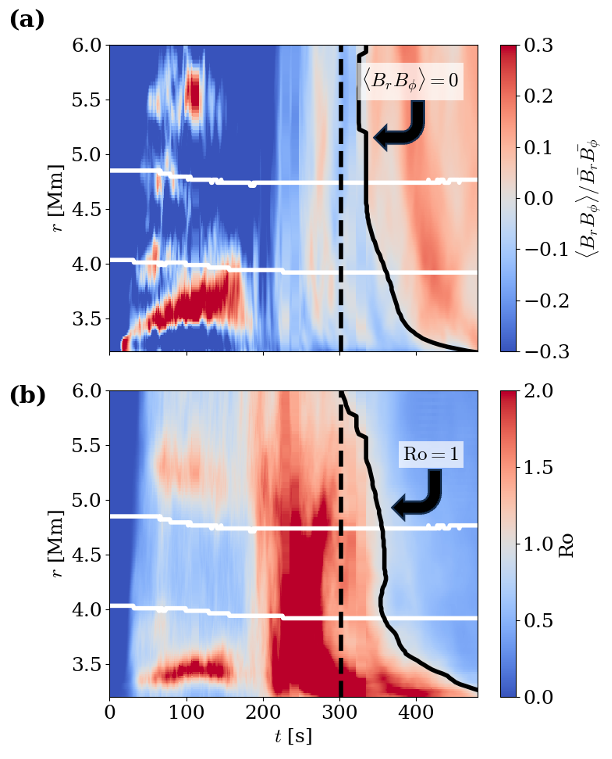}
 \caption{Time–radius diagrams of (a) the correlation between $B_r$ and $B_\phi$ (Eq.~\ref{eq:MS_Cor}), and (b) the Rossby number Ro, are shown using a cool-to-warm color scale (color bars are shown on the right). 
 The vertical axis indicates radius, and the horizontal axis denotes time.
 The O shell (the region between two white lines) reaches the quasi-steady states in the convective dynamo at $\approx 300$\,s (black dashed line). The black solid line in panel (a) indicates the region where the correlation changes its sign. The black solid line in panel (b) displays the region where Rossby number crosses unity.}
 \label{fig:Cor_Ro}
\end{figure}

Fig.~\ref{fig:Ro_def} (a)
illustrates the temporal evolution of rotation rates $\Omega/2\pi$ in mHz and convective velocity in cm/s in the MHD model of \citet{Varma_2023}, together with the corresponding variation in the Rossby number (Eq.~\ref{eq:Def_Ro}) within the O shell. Hereafter, we refer to the Favre-averaged rotation rate $\tilde{\Omega}_z$ as $\Omega$. The location of the O shell is defined later by Eq.~(\ref{eq:SA_Mod}) and indicated by white lines in Fig.~\ref{fig:Cor_Ro}.  
The convective velocity $v_\mathrm{conv}$ is calculated by the root mean square value of the fluctuating component of the Favre decomposition $v''$ (Appendix \ref{sec:appendix A}).
\\

Both the rotation rate and convective velocity exhibit a rapid decline around $t \approx 180$\,s (Fig.~\ref{fig:Ro_def}\,a).  
This corresponds to the end of the exponential growth phase of the magnetic field in the MHD simulation \citep[See Fig.~2 in][]{Varma_2023}.  
After this phase, the convective velocity continues to decrease.
On the other hand, the rotation rate reaches a minimum value of $1.1$\,mHz near $t \approx 350$\,s and subsequently begins to increase gradually, shown in the inset of Fig.~\ref{fig:Ro_def}.
This gradual increase corresponds to a 25\% rise over 100 seconds ($\sim$5 convective turnover times).
The initial decrease in rotation rate from $t = 100$ to $200$\,s, driven by the growth of magnetic fields, is consistent with the 1D AM transport model based on the Taylor-Spruit dynamo (Section~\ref{subsec:AM_mag}). While the dynamo type differs, both frameworks predict spin-down due to magnetic AM transport.
The later gradual increase after $t=350$\,s is a new feature identified for the first time in this model \citep[see also Fig.~9 in][]{Varma_2023}.
In addition, their MHD model shows persistent near-rigid rotation.
\\

Fig.~\ref{fig:Ro_def} (b) illustrates the temporal evolution of the convective Rossby number of the O shell.
The pressure scale height $H_p$ is adopted as the characteristic length scale $L$ in the definition of the Rossby number. The Rossby number increases rapidly around $t \approx 180$\,s.  
This rapid increase reflects the decrease of $\Omega$ by the efficient extraction of AM from the O shell due to the magnetic fields.  
In contrast, the Rossby number starts to decrease around $t \approx 250$\,s.  
This later decrease arises primarily from the continued decline in the convective velocity, although the mild increase of rotation rate also marginally decreases the Rossby number (Fig.~\ref{fig:Ro_def}\,a).
\\

\subsection{Definition of oxygen shell and suitable time for our analysis}\label{subsec:setup}

Here we provide a definition of the O shell used in this work.
The O shell corresponds to the convectively unstable layer generated by O-burning.
In the context of stellar evolution, such regions are commonly identified using the superadiabaticity via Eq.~(\ref{eq:SA}).
Although the superadiabaticity is expected to be positive in convectively unstable layers, we find that in these 3D simulations it attains slightly negative values even within regions regarded as the convective O shell in the progenitor provided by stellar evolution models. Equivalently, throughout the O shell there is a mildly positive, rather than flat radial entropy gradient.
Since this gradient is substantially weaker than those of the surrounding layers, we adopt a revised definition of the O shell to capture these regions, as follows:
\begin{eqnarray}
\displaystyle \mathrm{SA} + \mathrm{SA}_{\mathrm{mod}} > 0, \label{eq:SA_Mod}
\end{eqnarray}
where $\mathrm{SA}_{\mathrm{mod}}$ denotes an offset parameter introduced to ensure that the convective region is consistently identified by Eq.~(\ref{eq:SA_Mod}).
For the present analysis, we adopt $\mathrm{SA}_{\mathrm{mod}} = 0.3$, which yields a shell base consistent with the radius of peak O-burning energy generation.
In Fig.~\ref{fig:Cor_Ro}, the location of the O shell is identified as the region between the two white horizontal lines (4,000 – 4,700\,km).
Although the origin of the marginal convective stability ($\mathrm{SA}_{\mathrm{mod}} = 0.3$) of the O shell lies beyond the scope of this paper, a reduced efficiency of entropy mixing by convection compared to the MLT would produce this marginal convective stability.
The definition of the O shell via superadiabaticity provides a consistent framework for identifying the convective region in the 3D MHD simulation and forms the basis for the subsequent analyses.
\\

In \citet{Varma_2023}, transient behavior—such as the sudden decrease in rotation rate arising from the linear growth of convection and magnetic fields due to the initial condition is seen during $t<200$\,s in Fig.~\ref{fig:Ro_def} (a).
To ensure that the convective dynamo operating in the O shell has reached a quasi-steady state, we identify the time interval that satisfies the following criterion:
\begin{eqnarray}
\displaystyle \frac{d \ln \mathrm{KE}}{dt} - \frac{d \ln \mathrm{ME}}{dt} > 0, \label{eq:QS}
\end{eqnarray}
where KE and ME denote the kinetic energy density ($\frac{1}{2}\rho v''^2$) and magnetic energy density ($\frac{1}{8\pi}B^2$), respectively.
In this regime, the convective power varies on a shorter time scale than the magnetic energy.
Consequently, the magnetic field adjusts passively to the convective motions rather than being influenced by transient amplification or decay.
This condition isolates a quasi-steady state in which the magnetic field responds passively to the convective dynamics \citep{Seta_2021}.
In Fig.~\ref{fig:Cor_Ro}, the black dashed line shows the timing after which the system satisfies this quasi-steady condition.
In the O shell, this state is achieved after approximately $t \approx 300$\,s.
In this time range, the rotation rate decreases between 300 – 350\,s and begins to increase thereafter (Fig.~\ref{fig:Ro_def}\,a).
This change results from the AM transport under the convective dynamo in quasi-steady state.
This selection of time range ($300 - 480$\,s) ensures that the derived properties of AM transport represent quasi-steady behavior rather than transient effects. 
This interval spans approximately 12 convective turnover times.
In the following analysis, we focus on the AM transport occurring during this phase.
\\

\subsection{Relating the direction of angular momentum flux by Maxwell stress with the Rossby number}\label{subsec:dirct_MS}
From our analysis of AM evolution in Eq.~(\ref{eq:AM_Con}), we find that once the system reaches a quasi-steady state at $t \approx 300$\,s, the Maxwell stress becomes the dominant contributor to the AM transport, surpassing the other terms.
This dominance indicates that magnetic stresses largely govern the subsequent evolution of the rotation rate in the simulation.
Therefore, in the following analysis, we aim to incorporate the AM flux associated with the Maxwell stress into a framework suitable for 1D stellar evolution modeling.
We estimate the sign of the Maxwell stress based on the correlation between its components (e.g., Fig.~\ref{fig:Cor_Ro}\,a), and assess its magnitude later in Section~\ref{subsec:results_amplitudes}. 
These results provide a foundation for parameterizing magnetic angular momentum transport in 1D models.
\\

As shown in Eq.~(\ref{eq:AM_Con}), the direction of the AM flux due to the Maxwell stress is determined by the correlation between $B_r$ and $B_\phi$ via the definition of the spherical average.
To quantify the variation of this correlation in the simulation, we introduce the following diagnostic quantity:
\begin{eqnarray}
\displaystyle \mathrm{Cor}=\braket{B_rB_\phi}/\bar{B}_r\bar{B}_\phi. \label{eq:MS_Cor}
\end{eqnarray}
Here, a positive (negative) value indicates inward (outward) AM transport by the Maxwell stress at a given radius.
To calculate Cor in the O shell shown in Fig.~\ref{fig:Ro_def} (b), we take the radial average of Cor over the O shell defined by Eq.~(\ref{eq:SA_Mod}).
Fig.~\ref{fig:Ro_def}\,(b) shows the time evolution of this correlation (orange solid line) within the O shell.
The correlation changes sign from negative to positive around $t \approx 350$\,s. This indicates the transition of magnetic AM transport from outward to inward. We also note that the absolute value of Cor in the region of interest is typically about 0.1, a result which we will use later in Section~\ref{sec:model}.
We note that a standard deviation of Cor is 0.04 when restricting the analysis to time intervals with a fixed sign of $\mathrm{Cor}$ in the quasi-steady state.
\\

To characterize a potential relationship between the correlation $\mathrm{Cor}$ and the Rossby number $\mathrm{Ro}$, we examine the spatial and temporal evolution of these quantities in the O shell, as shown in Fig.~\ref{fig:Cor_Ro}. The white solid lines marks the location of the O shell boundaries, while the black dashed line indicates the time when the system reaches the quasi-steady state of the convective dynamo. During our timescale of interest ($\gtrsim 300$\,s), we now investigate a potential relationship between the direction of AM flux (via Cor) and Ro. When $\mathrm{Ro} > 1$ (red regions in Fig.~\ref{fig:Cor_Ro}\,b, $t \lesssim 350$\,s), $\mathrm{Cor}$ is negative (blue regions in Fig.~\ref{fig:Cor_Ro}\,a), indicating that the Maxwell stresses transport AM outward. In contrast, when $\mathrm{Ro} < 1$ (blue regions in Fig.~\ref{fig:Cor_Ro}\,b, $t \gtrsim 350$\,s), $\mathrm{Cor}$ becomes positive (red regions in Fig.~\ref{fig:Cor_Ro}\,a), showing that the direction of magnetic AM transport reverses and moves inward. The black solid line in Fig.~\ref{fig:Cor_Ro} (b) marks the $\mathrm{Ro} = 1$ boundary. Notably, the spatial and time location where the transitions from $\mathrm{Ro} > 1$ to $\mathrm{Ro} < 1$ closely coincides with the position where $\mathrm{Cor}$ changes sign from negative to positive (black solid line in Fig.~\ref{fig:Cor_Ro}\,a).
\\

This spatial and temporal agreement of the two transitions (solid black lines in Figs.~\ref{fig:Cor_Ro}\,a~and~b) indicates that the direction of magnetic AM transport is governed by the Rossby number.
When $\mathrm{Ro}>1$ (before $t \simeq 340$\,s), correlations are negative corresponding to outward AM transport. 
This is consistent with the trend reported in solar convection zone simulations \citep{Hotta_2022}. 
As the system evolves and the Rossby number decreases below unity ($\mathrm{Ro}<1$, when $t \gtrsim 350$\,s), the correlation becomes positive. This corresponds to inward transport. Such inward AM transport driven by magnetic fields has not been considered in stellar evolution models.

\subsection{Amplitude of angular momentum flux by Maxwell stress}\label{subsec:results_amplitudes}
Having examined the dependence of the direction of AM transport on the Rossby number in Section~\ref{subsec:dirct_MS}, we now formulate the amplitude of the AM flux.  
As shown in Eq.~(\ref{eq:AM_Con}), the AM flux due to the Maxwell stress contribution to AM transport scales with $\braket{B_rB_\phi}$, and its sign is determined by the correlation between $B_r$ and $B_\phi$ (Eq.~\ref{eq:MS_Cor}).  
We rewrite the Maxwell stress component in Eq.~(\ref{eq:AM_Con}) in terms of Cor as follows:
\begin{eqnarray}
\displaystyle -\frac{1}{4\pi}\Braket{B_rB_\phi r\sin \theta}
&&\approx -\frac{r}{2}\frac{\Braket{B_rB_\phi}}{\bar{B_r}\bar{B_\phi}}\frac{\bar{B_r}\bar{B_\phi}}{4\pi}\nonumber\\
&&\approx -\frac{r}{2}\mathrm{Cor}\,\mathrm{ME}.
\label{eq:maxwell_amp_lead}
\end{eqnarray}
In this expression, we approximate $\bar{B_r}\bar{B_\phi}/4\pi$ by the magnetic energy $\mathrm{ME}= \bar{B}^2/8\pi$.
Thus, the amplitude of the Maxwell-stress term is set by $\mathrm{ME}$.
We now investigate the Rossby number dependence of $\mathrm{ME}$.
\\

Observations, analytical studies, and numerical simulations of low mass stars have demonstrated that the magnetic energy generated by convective dynamos depends on the Rossby number \citep{See_2019,Reiners_2022,Augustson_2019,Brun_2022}.  
In particular, \citet{Augustson_2019} derived the following scaling relation based on a balance between the Coriolis force, Lorentz force, and inertia.  
This relation has been shown to reproduce results from a wide range of dynamo simulations, including those of solar-type stars and the convective cores of massive main-sequence stars.  
The scaling law expresses the ratio of magnetic energy ($\mathrm{ME}=\bar{B}^2/8\pi$) to kinetic energy ($\mathrm{KE}=\rho{v''}^2/2$) as:
\begin{eqnarray}
\displaystyle \mathrm{ME}/\mathrm{KE}=a\,\mathrm{Ro}^{-1}+b, \label{eq:MEKE_Ro}
\end{eqnarray}
where $a$ and $b$ are coefficients that depend on the efficiency of magnetic energy generation.
In Figs.~\ref{fig:Ro_def}~and~\ref{fig:Cor_Ro}, we have characterized the temporal evolution of the Rossby number.  
By computing $\mathrm{ME}/\mathrm{KE}$, we now examine whether this scaling law holds in O shell of the 3D MHD simulation.  
Note that the scaling law in \citet{Augustson_2019} is further normalized by the magnetic Prandtl number ($\mathrm{Pm}$). 
In contrast, the magnetic and velocity fields in \citet{Varma_2023} do not include explicit dissipation; instead, a similar artificial viscosity scheme is applied to both quantities hence $\mathrm{Pm}$ is assumed to be unity \citep{Lesaffre_2007}.
In addition, the dependence on $\mathrm{Pm}$ cannot be systematically examined in the present study. 
For these reasons, we adopt $\mathrm{Pm}=1$ in our analysis and employ a scaling law that does not explicitly include the effect of the magnetic Prandtl number (Eq.~\ref{eq:MEKE_Ro}).
\\

\begin{figure}
 \includegraphics[width=\columnwidth]{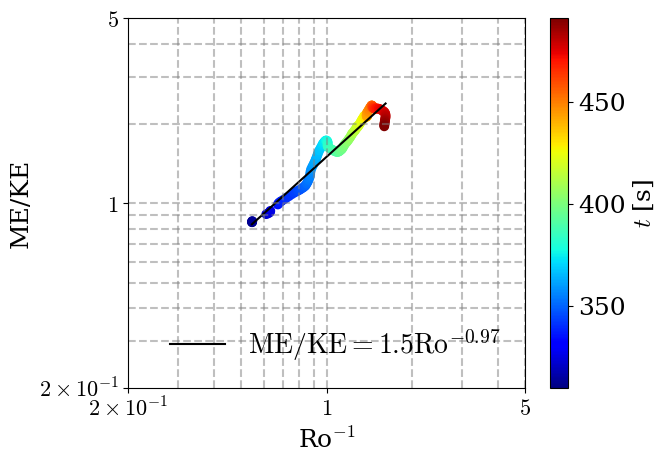}
 \caption{The horizontal axis shows the inverse of the Rossby number, and the vertical axis shows the ratio between the magnetic and kinetic energy.
The color of each point indicates the corresponding time in the simulation that satisfies the quasi-steady state condition by the rainbow colorbar ($t =$ 300 – 480\,s; Eq.~\ref{eq:QS}).
The black solid line represents the best fit to the quantities taken from the 3D MHD simulation.}
 \label{fig:ME_Ro}
\end{figure}

In Fig.~\ref{fig:ME_Ro}, we plot the ratio between magnetic and kinetic energies ME/KE (vertical axis) against the inverse Rossby number (horizontal axis) on a log-log scale.
The color indicates the time, limited to our timescale of interest. 
The black solid line denotes the best-fit scaling relation derived from our data. The best-fit power-law index of $-0.97 \pm 0.02$, obtained using a nonlinear least-squares fitting procedure, agrees well with the scaling proposed by \citet{Augustson_2019}.
The coefficients in Eq.~\ref{eq:MEKE_Ro} are fitted to $a=1.5\pm0.1$ and $b=0.083\pm0.08$ respectively.
As the Rossby number decreases, the ratio $\mathrm{ME}/\mathrm{KE}$ increases, showing that rapid rotation enhances the efficiency of magnetic energy generation. This confirms that the convective dynamo in the O shell obeys the same rotational dependence observed in the low mass stars. Accordingly, the magnetic energy (ME) in the O shell is well described by the kinetic energy (KE) and the Rossby number (Ro), establishing a quantitative link between magnetic field strength and rotational dynamics.
\\

We now seek to formulate the KE of the O shell by the quantities given by 1D stellar evolution models. We first express the convective velocity in terms of the energy transported by convection. This expression has been validated both analytically \citep[e.g.,][]{Arnett_2009} and through 3D HD simulations \citep{Muller_2016}, and is given by
\begin{eqnarray}
\displaystyle v_{\mathrm{conv}}^{3} \propto \int_{0}^{r} \dot{Q} dV, \label{eq:HD_vconv}
\end{eqnarray}
where $\dot{Q}$ denotes the nuclear energy generation rate.
Here, $v_{\rm conv}$ is a function of radius, and the volume integral on the right-hand side is taken over the spherical region enclosed within the radius at which $v_{\rm conv}(r)$ is evaluated.
This relation indicates that the convective velocity is determined by the nuclear energy released beneath the convective region.
Therefore, variations in nuclear burning change the strength of convective motion and, consequently, the kinetic energy within the shell.
\\

However, ME exceeds the convective KE in our simulation over the timescale of interest (Fig.~\ref{fig:ME_Ro}). Therefore, $v_\mathrm{conv}$ based on convective energy transport via KE alone is not appropriate. Instead we assume that the nuclear energy released in the shell is divided into two components: convective energy flux and Poynting flux. Here, the fraction distributed to each component is determined by the ratio of kinetic to magnetic energy. Based on this assumption, we predict that the convective velocity can be scaled by the portion of the energy flux that goes into the convective energy flux, as follows:
\begin{eqnarray}
\displaystyle v^3_{\mathrm{conv}}\propto \frac{1}{1+\mathrm{ME}/\mathrm{KE}}\int_{0}^{r} \dot{Q} dV. \label{eq:KE_Qdot}
\end{eqnarray}
\begin{figure}
 \includegraphics[width=\columnwidth]{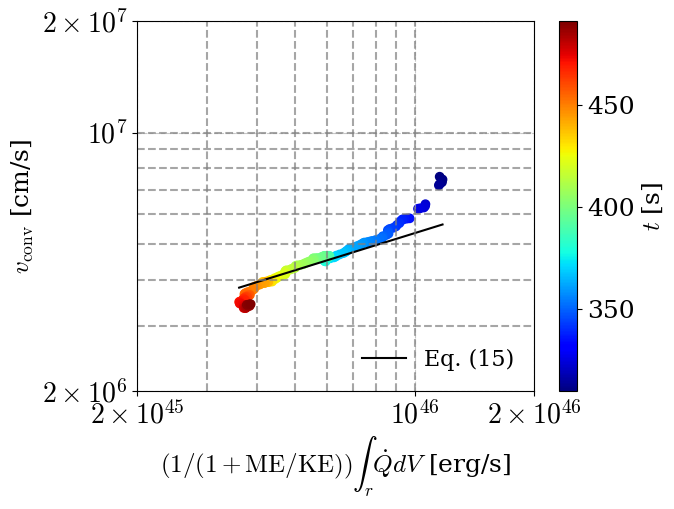}
 \caption{Comparison of our predicted convective energy flux based on the energy partition between convective and Poynting fluxes (Eq.~\ref{eq:KE_Qdot}) with simulation data. The convective velocity $v_\mathrm{conv}$ in the O shell is shown on the vertical axis against the right-hand side of Eq.~(\ref{eq:KE_Qdot}) on a log-log scale. The color of each point indicates the corresponding time in the simulation during the quasi-steady-state phase ($t =$ 300 – 480\,s).
The black line represents a power-law with an index of 1/3.}
 \label{fig:v_conv}
\end{figure}
This expression links the convective velocity, which determines KE, to the nuclear burning rate, and the ratio (ME/KE).
Here, as in Eq.~(\ref{eq:HD_vconv}), $v_{\rm conv}$ is a function of radius, and the volume integral on the right-hand side is taken over the spherical region enclosed within the radius at which $v_{\rm conv}(r)$ is evaluated. In Fig.~\ref{fig:v_conv}, the convective velocity in the O shell (vertical axis) is shown against the right-hand side of Eq.~(\ref{eq:KE_Qdot}) (horizontal axis) on a log-log scale.
We compute the convective velocity $v_{\rm conv}(r)$ at each radius within the O shell—defined as the root-mean-square value of the fluctuating velocity component $v''$ obtained from the Favre decomposition—as well as the right-hand side of Eq.~(\ref{eq:KE_Qdot}), and show their respective root-mean-square values. The black line represents the $1/3$ power-law relation predicted by Eq.~(\ref{eq:KE_Qdot}). 
By incorporating the ratio between magnetic and kinetic energies (ME/KE) into Eq.~(\ref{eq:KE_Qdot}), the convective velocity is well described by the nuclear energy generation rate. 
This result confirms that our scaling successfully captures the influence of magnetic energy on convection.
Furthermore, this formulation naturally incorporates the magnetic suppression of the convective velocity from hydrodynamic convection (Eq.~\ref{eq:HD_vconv}).
A nonlinear least-squares fit to the data yields a power-law index of $0.37 \pm 0.01$, which is slightly steeper than the $1/3$ scaling. This deviation is primarily driven by data from the early phases of the quasi-steady state. As shown in Fig.~\ref{fig:v_conv}, the scaling progressively approaches the $1/3$ power-law as the system evolves.

In this Section, we have shown that the amplitude of the Maxwell stress can be estimated from the magnetic energy (ME) in Eq.~(\ref{eq:maxwell_amp_lead}), while Fig.~\ref{fig:ME_Ro} demonstrates that the ratio ME/KE of magnetic to kinetic energy depends on the Rossby number. Fig.~\ref{fig:v_conv} shows the dependence of the convective velocity on ME/KE and the nuclear energy generation rate $\dot{Q}$. Combining these results, we infer that ME is ultimately determined by the Rossby number and $\dot{Q}$. In the next Section, we formulate the amplitude of AM transport by the Maxwell stress based on these dependences.

\section{Model of Angular Momentum Transport by Magnetic Fields}\label{sec:model}
\subsection{Model for the Direction of Angular Momentum Flux} 
Based on the strong correlation between the Maxwell stress and the Rossby number (Fig.~\ref{fig:Cor_Ro}), we approximate the correlation of the Maxwell stress ($\mathrm{Cor}$) as a function of the Rossby number:
\begin{eqnarray}
\mathrm{Cor}=\frac{\braket{B_rB_\phi}}{\bar{B_r}\bar{B_\phi}}=f(\mathrm{Ro})=
  \begin{cases}
    \alpha  & \text{for $\mathrm{Ro}<\beta$}, \\
    -\alpha & \text{for $\mathrm{Ro}>\beta$}  \end{cases}.
\label{eq:def_f}
\end{eqnarray}
$\alpha$ represents the amplitude of the correlation, and $\beta$ denotes the critical Rossby number at which the correlation reverses sign. 
From the simulation during the quasi-steady phase, Fig.~\ref{fig:Cor_Ro} (a) shows that during our timescale of interest, the magnitude of Cor is typically 0.1. We therefore take $\alpha = 0.1$ from the solid orange line in Fig.~\ref{fig:Ro_def} (b). We take $\beta = 1$ from the solid black lines in Fig.~\ref{fig:Cor_Ro} (b).

\subsection{Model for the Amplitude of Angular Momentum Flux} 
The amplitude of the AM flux driven by the Maxwell stress can be estimated from the magnetic energy, shown by Eq.~(\ref{eq:maxwell_amp_lead}). Therefore, in this section, we model the magnetic energy (ME) itself. Following the results shown in Section~\ref{subsec:results_amplitudes} (Figs.~\ref{fig:ME_Ro}~and~\ref{fig:v_conv}), we derive ME as a product of two components: the ratio between the magnetic and kinetic energies ($\mathrm{ME}/\mathrm{KE}$), and the KE which is determined by $v_{\mathrm{conv}}$. Based on the clear dependence of $\mathrm{ME}/\mathrm{KE}$ on the Rossby number, and consistency with the theoretical scaling from convective dynamos in low-mass stars \citep{Augustson_2019}, we model
\begin{equation}
\frac{\mathrm{ME}}{\mathrm{KE}} \equiv g(\mathrm{Ro}) = a_{\mathrm{sim}}\,\mathrm{Ro}^{-1} + b_{\mathrm{sim}},
\label{eq:def_g}
\end{equation}
where $a_{\mathrm{sim}}$ and $b_{\mathrm{sim}}$ are fitting parameters obtained from the simulation during the quasi-steady phase. 
We obtain $a_{\mathrm{sim}} = 1.5\pm{0.1}$ and $b_{\mathrm{sim}} = 0.083\pm0.08$.
Furthermore, the analysis of the energy flux (Fig.~\ref{fig:v_conv}) indicates that 
the convective velocity is determined by the fraction of the total energy flux that is carried by convection.
This relation can be modeled as:
\begin{eqnarray}
v_{\mathrm{conv}}
  &=&
  \left[
    \frac{2}{\rho}
    \frac{\mathrm{KE}}{\mathrm{KE} + \mathrm{ME}}
    \frac{\int_{0}^{r} \dot{Q}\,dV}{4\pi r^2}
  \right]^{1/3} \nonumber\\
  &=&
  \left[
    \frac{2}{\rho}
    \frac{1}{1 + g(\mathrm{Ro})}
    \frac{\int_{0}^{r} \dot{Q}\,dV}{4\pi r^2}
  \right]^{1/3},
\label{eq:vconv_Model}
\end{eqnarray}
where $\dot{Q}$ represents the nuclear energy generation rate, and $\rho$ is the density. Here ME/KE is replaced by $g(\mathrm{Ro})$ using Eq.~(\ref{eq:def_g}).
The magnetic energy (ME) can finally be expressed as:
\begin{eqnarray}
\mathrm{ME}
&=&
\frac{\mathrm{ME}}{\mathrm{KE}}\,\mathrm{KE}
\approx
\frac{\mathrm{ME}}{\mathrm{KE}}\,\frac{\rho}{2}\,v_{\mathrm{conv}}^2 \nonumber\\
&\approx&
g(\mathrm{Ro})\,\frac{\rho}{2}
\left[
  \frac{2}{\rho}
  \frac{1}{1 + g(\mathrm{Ro})}
  \frac{\int_{0}^{r} \dot{Q}\,dV}{4\pi r^2}
\right]^{2/3}.
\label{eq:ME_model}
\end{eqnarray}
\\

\subsection{Complete Model of the Maxwell Stress} 
We now present the complete formulation of the AM flux by the Maxwell stress, $-\frac{1}{4\pi}\braket{B_r B_\phi r \sin \theta}$. As discussed in Section~\ref{subsec:results_amplitudes}, 
the AM flux by the Maxwell stress can be decomposed into two components: the correlation coefficient $\mathrm{Cor}$, which determines its direction, and the magnetic energy (ME), which characterizes its amplitude (Eq.~\ref{eq:maxwell_amp_lead}). 
\\

The correlation $\mathrm{Cor}$ is modeled as a function of the Rossby number in Eq.~(\ref{eq:def_f}), 
while the magnetic energy ME is modeled as a function of density $\rho$, Rossby number ($\mathrm{Ro}$), 
and nuclear energy generation rate $\dot{Q}$ in Eq.~(\ref{eq:ME_model}). 
Finally inserting these models into Eq.~(\ref{eq:maxwell_amp_lead}), we obtain the following expression for the AM flux driven by the Maxwell stress,
\begin{eqnarray}
&&-\frac{1}{4\pi}\Braket{B_r B_\phi r \sin \theta}\nonumber\\
&&\approx
-\frac{r}{2}\,f(\mathrm{Ro})\,g(\mathrm{Ro})\,\frac{\rho}{2}
\left[
  \frac{2}{\rho}
  \frac{1}{1 + g(\mathrm{Ro})}
  \frac{\int_{0}^{r} \dot{Q}\, dV}{4\pi r^2}
\right]^{2/3}.\nonumber\\
\label{eq:MS_Model}
\end{eqnarray}
This model estimates the Maxwell stress using the Rossby number $\mathrm{Ro}$, the density $\rho$, and the energy generation rate $\dot{Q}$.
All of these quantities can be directly obtained from 1D stellar evolution calculations.
In practice, the Rossby number can be derived from the local angular velocity and convective turnover timescale, which are commonly available outputs in 1D models. 
The quantities
$\rho$ and $\dot{Q}$ are fundamental outputs of 1D stellar evolution calculations.
Accordingly, the present formulation enables computation of the AM flux by the Maxwell stress in the convective zone at each time step of a stellar evolution calculation. 
Using the radial profiles of $\rho$, $\mathrm{Ro}$, and $\dot{Q}$, the corresponding radial profile of the Maxwell stress can be obtained.
This can be implemented, in principle, into 1D stellar evolution calculations to model AM transport by magnetic fields in convection zones.

\subsection{Comparison with 3D simulation} 

\begin{figure}
 \includegraphics[width=\columnwidth]{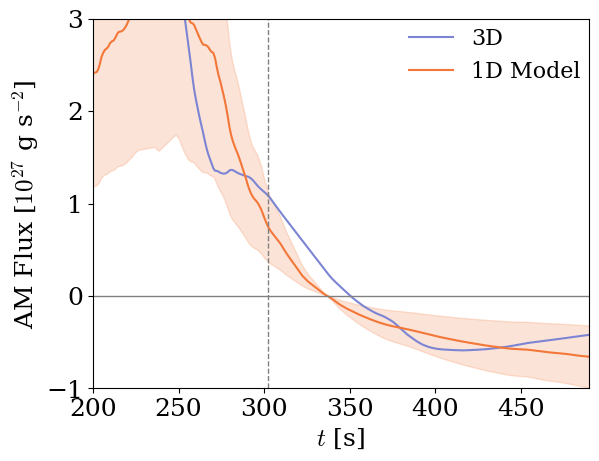}
 \caption{The temporal evolution of the angular momentum flux due to the Maxwell stress (Eq.~\ref{eq:AM_Con}) in the 3D MHD simulation (blue line). Our 1D model, given by Eq.~(\ref{eq:MS_Model}) is shown by the orange line.
The orange shaded region represents the uncertainty of the 1D model, accounting for the combined standard deviation of the model parameters.}
The flux is measured at the middle of the O shell ($4,500$\,km).
The gray dashed line indicates the time at which the convective dynamo in the O shell reaches the quasi-steady state.
After 340\,s and 350\,s for the 1D and 3D models, respectively, the angular momentum is transported inward, coinciding with the timing when the rotation rate begins to spin up (see Fig.~\ref{fig:Ro_def} a).
 \label{fig:AM_Model}
\end{figure} 

In Fig.~\ref{fig:AM_Model}, the temporal evolution of the AM flux associated with the Maxwell stress (Eq.~\ref{eq:AM_Con}) computed directly from the 3D MHD simulation is shown by the blue line. The estimate using our 1D model in Eq.~(\ref{eq:MS_Model}) is shown by the orange line, where the spherically averaged density $\rho$, Rossby number $\mathrm{Ro}$, and energy generation rate $\dot{Q}$ are taken from the each timestep of the 3D MHD simulation. 
The orange shaded region indicates the uncertainty of the 1D model, corresponding to the combined standard deviation of the model parameters. This accounts for the standard deviation of $\mathrm{Cor}$ ($\sigma_{\mathrm{Cor}} = 0.04$) in Eq.~(\ref{eq:def_f}), the standard deviations of the fitted coefficients $a_{\rm sim}$ and $b_{\rm sim}$ in Eq.~(\ref{eq:def_g}), and the difference between the assumed $1/3$ exponent and the fitted value of $0.37$ in Eq.~(\ref{eq:vconv_Model}).
The gray dashed line marks the time ($t = 300$\,s) from which the convective dynamo in the O shell satisfies a quasi-steady state, as defined by Eq.~(\ref{eq:QS}). 
The primary objective of this comparison is to examine whether our 1D model can reproduce the AM transport by the Maxwell stress during the quasi-steady phase.
Therefore, the interval after $t = 300$\,s serves as a key benchmark for validating the applicability of our 1D formulation.
\\

In Fig.~\ref{fig:AM_Model}, the direction of the AM flux reverses from outward to inward transport in the 3D model at 350\,s. 
In the 1D model, the reversal occurs at 340\,s.
This transition coincides with the end of spin down, and the onset of the spin-up phase. 
This is when the rotation rate begins to increase as shown in the inset of Fig.~\ref{fig:Ro_def} (a). The 1D model reproduces this reversal remarkably well. Moreover, the magnitude of the AM flux predicted by the 1D model closely matches that obtained from the 3D simulation during the quasi-steady phase ($t > 300$\,s).
These results demonstrate that our 1D model from Eq.~(\ref{eq:MS_Model}) successfully captures both the direction and amplitude of the Maxwell stress associated with the convective dynamo in the O shell.

Moreover, when the uncertainty of the 1D model is taken into account, the AM flux in the 3D model is reproduced by the 1D formulation for most of the quasi-steady phase ($t > 300$\,s). We note that the dominant contribution to the uncertainty of the 1D model arises from the standard deviation of $\mathrm{Cor}$. Deviations between the 1D and 3D models are most noticeable around the transition time near $t \simeq 340$\,s, which can be attributed to the simplified treatment in the 1D model: as described in Eq.~(\ref{eq:def_f}), the sign of $\mathrm{Cor}$ changes instantaneously once the Rossby number drops below unity. In contrast, the AM flux and $\mathrm{Cor}$ in the 3D simulation adjust to changes in the Rossby number over approximately one convective turnover time. This short temporal offset is not expected to be significant for practical stellar evolution calculations.

\section{Discussion}\label{sec:Discussions}
We discuss the implications and limitations of our 1D model for AM transport by Maxwell stress, given by Eq.~(\ref{eq:MS_Model}).
First, we consider potential applications over stellar evolution timescales.

\subsection{Angular momentum transport in inner regions} \label{subsec:AM_inner}

Historically, AM transport by magnetic fields has been modeled as an outward flux, in which AM is transferred from the inner to the outer regions.
This leads to a net spin-down over stellar evolution timescales \citep[e.g.,][]{Suijs_2008,Heger_2005}.
Magnetic fields have therefore been regarded as a promising mechanism to reconcile the long-standing discrepancy between the observed and theoretically predicted core rotation rates \citep[e.g.,][]{Hartogh_2019}. 
\\

However, in the 3D MHD simulation of \citet{Varma_2023}, magnetic fields do not necessarily transport AM outward. Fig.~\ref{fig:Ro_def} (a) shows an increase in the rotation rate within the O shell around $t \simeq 350$\,s. 
This spin-up of the oxygen shell at 350\,s can also be seen in Fig.~9 of \citet{Varma_2023}.
We show in Fig.~\ref{fig:Cor_Ro} that the Rossby number determines the correlation between $B_r$ and $B_\phi$, which then governs the magnetic AM flux (Eq.~\ref{eq:MS_Cor}). 
By expressing the Maxwell stress in terms of this correlation, our formulation (Eq.~\ref{eq:MS_Model}) indeed predicts negative (inward) AM transport into the O shell (4,500\,km) after $t \simeq 340$\,s (Fig.~\ref{fig:AM_Model}). 
As expected, the inner oxygen shell region begins to spin up at nearly the same time ($t \simeq 350$\,s; Fig.~\ref{fig:Ro_def}\ a), a consequence of the onset of the inward Maxwell stress AM flux predicted by our 1D model (Fig.~\ref{fig:AM_Model}).
\\

Our 1D AM transport model is therefore the first to account for inward AM transport by magnetic fields. In the context of stellar evolution, the Maxwell stress drives inward AM transport when the Rossby number is below unity ($\mathrm{Ro}<1$) in convectively unstable regions. This regime corresponds to a condition in which magnetic spin-down is suppressed. As a result, the AM flux instead contributes to the spin-up of the stellar core during this oxygen shell burning simulation.
This transition from spin-down to spin-up should not be interpreted as a simple readjustment following the initial spin-down phase.
Instead, in our interpretation it is driven by the reduction of the nuclear energy generation rate $\dot{Q}$, which leads to a decrease in the Rossby number and thereby triggers inward angular momentum transport and core spin-up. This change in $\dot{Q}$ occurs due the suppression of mixing in the convective shell due to the strong magnetic fields \citep{Varma_2023}. We therefore expect this spin-up to persist as long as $\dot{Q}$ does not increase due to other evolutionary effects.
Establishing a broader applicability will require further simulations, e.g., testing against a complete shell burning simulation \citep{Rizzuti_2023} in MHD, which we leave to future work. 
Note that in the present simulation the Ne and C shells have not yet reached a quasi-steady state and are therefore not suitable for a robust verification of the prescription.
\\

Based on our 1D angular momentum (AM) transport model in convective regions, one may anticipate that combining it with magnetic AM transport models in radiative zones, such as that proposed by \citet{Fuller_2019}, could result in a less efficient spin-down of the stellar core than previously expected. 
However, our model is sensitive to the Rossby number and to the nuclear energy generation rate, both of which vary significantly over the course of stellar evolution. This sensitivity makes a quantitative assessment of the combined impact on the global AM evolution challenging at this stage. In order to robustly evaluate the resulting core spin evolution and its compatibility with observational constraints, it will therefore be necessary to incorporate our transport prescription into full 1D stellar evolution calculations.
\\

We can, however, speculate on the applicability to 1D stellar evolution models over several burning phases. 
We first note that our 1D model contains several empirical parameters that must be calibrated depending on the stellar type and evolutionary stage considered, i.e., $\alpha$, $\beta$ in Eq.~(\ref{eq:def_f}), and $a_{\mathrm{sim}}$, $b_{\mathrm{sim}}$ in Eq.~(\ref{eq:def_g}). The parameter $\alpha$, which characterizes the magnitude of the correlation between $B_r$ and $B_\phi$, depends on the properties of turbulence and likely varies with the Reynolds number, i.e., the numerical resolution of the simulations. Here, we note that our 1D model from (Eq.~\ref{eq:MS_Model}) underestimates the Maxwell stress during the early phase prior to $200$\,s. This discrepancy can be attributed to the unusually strong correlation between $B_r$ and $B_\phi$ ($|\alpha|\approx0.5$) at that time.
The correlation $\alpha$ during this early phase is likely sensitive to the initial magnetic field configuration.
The coefficients $a_{\mathrm{sim}}$, and $b_{\mathrm{sim}}$, which determine the scaling of $\mathrm{ME}/\mathrm{KE}$, should ideally be constrained through short-term direct numerical simulations tailored to each stellar structure and evolutionary phase.
\\

The energy generation rate $\dot{Q}$ can taken from 1D stellar evolution calculations when applying our model.
It is also important that these $\dot{Q}$ from 1D stellar evolution models \citep[e.g.,][]{Paxton_2013} appropriately reflect the effects of 3D convection and magnetic fields.
In particular, chemical mixing processes can strongly influence $\dot{Q}$. While Eq.~(\ref{eq:MS_Model}) can in principle predict fluxes, we caution that this suppression of mixing by magnetic fields should be incorporated into 1D simulations before considering magnetic AM transport. 
While Eq.~(\ref{eq:MS_Model}) can in principle predict fluxes, we caution that the suppression of chemical mixing by magnetic fields should be incorporated into 1D stellar evolution models before considering magnetic AM transport. Such suppression has also been reported in 3D MHD simulations of O shell by \citet{Leidi_2023}.
Developing a 1D prescription for chemical mixing processes in magnetoconvection constitutes one of the directions for our future work.

\subsection{Convective dynamo properties}\label{subsec:Convective dynamo}
Finally, we discuss an interesting feature identified in the 3D MHD model related to the convective dynamo.
Although this topic lies beyond the primary scope of this study, we find evidence for the operation of a dynamo mechanism within the O shell.
In Fig.~\ref{fig:Cor_Ro}\,(a), a characteristic pattern appears in which a positive correlation Cor (in red) propagates from the upper carbon shell at 8,000\,km to the lower regions of the oxygen shell at 4,000\,km around $t \approx 400$\,s.
This behavior can be interpreted as the propagation of a dynamo wave, consistent with an $\alpha$-$\Omega$-type dynamo \citep{Parker_1955, Yoshimura_1975}.
The presence of anti-solar differential rotation at this time, accompanied by positive (negative) kinetic helicity in the northern (southern) hemisphere, is consistent with the direction of propagation predicted by \citet{Yoshimura_1975}.
The identification of such a dynamo wave within the O shell provides strong evidence that the observed magnetic field amplification is driven by an internal convective dynamo, rather than by a Tayler–Spruit dynamo operating in the surrounding stably stratified layers. 
We leave a detailed comparison for future work.

\section{Conclusion}\label{sec:conclusion}
In massive stars, the rotation and magnetic field of the core strongly influence the subsequent core-collapse dynamics.
This motivates a careful treatment of angular-momentum (AM) evolution in stellar models. 
In practice, various recipes to treat the evolution of rotation rate in stellar interiors has been incorporated into one-dimensional (1D) stellar evolution frameworks.
Purely hydrodynamic (HD) models are known to overestimate core rotation by roughly two orders of magnitude \citep[e.g., red-giant cores; ][]{Moyano_2022}, but including magnetic AM transport in radiative zones via the Tayler–Spruit dynamo \citep{Spruit_2002} can substantially narrow this gap \citep{Fuller_2019}.
However, magnetic AM transport by convective dynamo has not yet been formulated.
\\

In recent 3D magnetohydrodynamic (MHD) simulations of convective burning shells in core-collapse progenitors \citep{Varma_2023}, magnetic fields generated by a convective dynamo strongly impact the rotation state of the oxygen-burning shell (O shell). 
The initial magnetic field, as small as $10^7$\,G, is amplified to $10^{11}$\,G within $t \approx 180$\,s.
After this amplification, the rotation rate drops by nearly an order of magnitude.
This indicates efficient magnetic AM transport by Maxwell stresses (Fig.~\ref{fig:Ro_def}\,a). Motivated by this, our goal is to develop a 1D prescription that captures the magnetic AM transport produced by convective dynamo. 
\\

In the present work, we took the 3D MHD simulation by \citet{Varma_2023} and analyzed its behavior in the time domain where the convective dynamo is in quasi-steady state at $t\geq$300\,s; in the O shell.
The O shell is identified by a superadiabaticity based criterion. Within this time  window, the rotation rate continues to decline until $t \approx 350$\,s and then reverses to a gradual spin-up. 
\\

Motivated by extensive 3D MHD studies of convective dynamos in solar-type stars, we focus our analysis on the Rossby number, which is the key control parameter for rotating convection. 
In solar-like convection, rotationally-induced anisotropy produces turbulent Reynolds stresses whose direction of AM transport switches with $\mathrm{Ro}$ \citep[e.g.,][]{Kitchatinov_1993,Miesch_2005}.
In MHD, the ideal ``frozen-in'' condition couples the magnetic field to the flow \citep{Alfven_1943}, so that the fluid velocity vector and magnetic field vector are closely aligned.
This alignment relates the Maxwell and Reynolds stresses, implying a corresponding Rossby dependence for the magnetic AM transport. On this basis, we decompose the AM flux by the Maxwell stress flux into (i) a direction set by the field correlation between $B_r$ and $B_\phi$ (Eq.~\ref{eq:MS_Cor}) and (ii) an amplitude set by the magnetic energy (Eq.~\ref{eq:maxwell_amp_lead}).
\\

We first compared the correlation $\mathrm{Cor}$ with the Rossby number. During the $\mathrm{Ro}>1$ phase we find that $\mathrm{Cor}<0$, indicating outward AM transport by the Maxwell stress. After the system transitions to $\mathrm{Ro}<1$, the correlation reverses to $\mathrm{Cor}>0$, implying inward transport. This correspondence shows that the direction of the magnetic AM flux is controlled by the Rossby number. Based on this result, we approximate the direction of flux by the function of Rossby number via Eq.~(\ref{eq:def_f}).
\\

Next, we analyzed the magnetic energy that sets the amplitude of the Maxwell stress. Rossby number dependence between magnetic and kinetic energies ME and KE has been pointed out by observations, analytic approaches, and simulations of solar-type stars.
During the quasi-steady phase, the ratio $\mathrm{ME/KE}$ in our simulation is well described by the Rossby-dependent scaling (Eq.~\ref{eq:MEKE_Ro}) suggested by \citet{Augustson_2019}.
For the convective velocity, we extended the standard scaling of convective velocity which considers the balance between the nuclear energy generation rate and convective velocity.
We also account for the partitioning of the total flux into convective and Poynting fluxes, yielding the modified scaling by Eq.~(\ref{eq:KE_Qdot}). The convective velocity in our simulation is well reproduced by the prediction formulated in Eq.~(\ref{eq:KE_Qdot}). Combining these ingredients, we constructed a model of ME as the amplitude of the Maxwell stress in terms of the Rossby number, the nuclear energy generation rate, and the density in Eq.~(\ref{eq:ME_model}).
\\

Finally, we combined these to obtain a complete 1D model of AM flux by the Maxwell stress in Eq.~(\ref{eq:MS_Model}). In this formulation, the flux is expressed solely in terms of quantities accessible from 1D stellar evolution models, namely, the Rossby number, the nuclear energy generation rate, and the density.
We validate our 1D model by comparing it with the actual Maxwell stress in the 3D MHD simulation. Input parameters for our 1D model are taken from the 3D MHD simulation. Our model reproduces the sign reversal of the angular momentum transport from outward to inward near $t \approx 350$\,s. It also matches the absolute magnitude of the flux during the quasi–steady phase after $\geq 300$\,s. Notably, the reversal time coincides with the onset of the spin–up in the 3D model.
\\

Our results have important implications for stellar evolution. 
Because both the sign and efficiency of magnetic AM transport vary with the Rossby number, magnetic spin-down of stellar cores would be suppressed near $\mathrm{Ro}\approx 1$.
Consequently, magnetic braking can be less effective than predicted by Rossby-number independent prescriptions, such as standard implementations of the Tayler–Spruit dynamo in the radiative zone \citep{Fuller_2019}. Implementing a Maxwell stress model dependent on Rossby number is thus essential in 1D MHD stellar evolution codes. 
\\

We also identified several limitations and caveats. 
The correlation coefficient between $B_r$ and $B_\phi$ (which sets the sign and magnitude of the Maxwell flux) is thought to depend on turbulent properties, i.e., Reynolds numbers and numerical resolution.
Furthermore, chemical mixing processes are expected to affect $\dot{Q}$ in our 1D model (Eq.~\ref{eq:MS_Model}). Since $\dot{Q}$ sets the amplitude of the AM flux, constructing a 1D prescription of chemical mixing informed by 3D magnetoconvection is an important next step and one of the directions for our future work.
\\

In summary, by linking Rossby number regulated correlations, dynamo scaling, and convective energetics, we develop a 1D magnetic AM transport model that captures key features of the 3D MHD behavior, including inward AM transport not represented in existing 1D prescriptions. 
The next critical step is to implement and test this prescription within stellar evolution codes (e.g., MESA) to determine how convective dynamos quantitatively alter core rotation over stellar evolution time scales. 

\section*{Data availability}
The data underlying this article will be shared on reasonable request
to the authors, subject to considerations of intellectual property law.

\acknowledgments
RS is supported by JST SPRING grant No. JPMJSP2110. LM acknowledges support from RIKEN iTHEMS and the Hakubi project at Kyoto University. 
KM is supported by JSPS KAKENHI grant No. JP24KK0070, JP24H01810, and JP24K00682. 
TY is supported by JSPS KAKENHI grant No. JP21H04492. 
BM acknowledges support from the Australian Research Council through Discovery Project DP240101786. 
We acknowledge computer time allocations from Astronomy Australia Limited's ASTAC scheme and the National Computational Merit Allocation Scheme (NCMAS). Some of this work was performed on the Gadi supercomputer with the assistance of resources and services from the National Computational Infrastructure (NCI), which is supported by the Australian Government.  Some of this work was performed on the OzSTAR national facility at Swinburne University of Technology. The OzSTAR program receives funding in part from the Astronomy National Collaborative Research Infrastructure Strategy (NCRIS) allocation provided by the Australian Government, and from the Victorian Higher Education State Investment Fund (VHESIF) provided by the Victorian Government.

\appendix

\section{Averaging procedures for turbulent analysis}\label{sec:appendix A}

In order to isolate the turbulent (fluctuating) components of the flow field in the present analysis, we employ the Favre (density-weighted) decomposition \citep{Favre_1965}, as in previous stellar convection studies \citep[e.g.,][]{McNeill_2020,McNeill_2022}.
All averages in this work are taken over spherical surfaces at a fixed radius unless stated otherwise.
For a generic field \(X(r,\theta,\phi,t)\) we define the spherical (solid-angle) average as:
\begin{equation}
\langle X \rangle(r,t) \equiv \frac{1}{4\pi}\int_{0}^{2\pi}\int_{0}^{\pi} X(r,\theta,\phi,t)\,\sin\theta\,d\theta\,d\phi .
\end{equation}
The Favre (mass-weighted) average of \(X\) is then given by:
\begin{equation}
\widetilde{X}(r,t) \equiv \frac{\langle \rho X \rangle(r,t)}{\langle \rho \rangle(r,t)} ,
\end{equation}
where \(\rho\) is the mass density.  The turbulent (fluctuating) component associated with the Favre decomposition is defined by:
\begin{equation}
X''(r,\theta,\phi,t) \equiv X(r,\theta,\phi,t) - \widetilde{X}(r,t).
\end{equation}
By definition, the fluctuation $X''$ satisfies the mass-weighted orthogonality condition,
\begin{equation}
\langle \rho\, X'' \rangle(r,t) = 0 .
\end{equation}
The root mean square amplitude of the turbulent field can be characterized in an area-weighted form as follows:
\begin{align}
\bar{X}(r,t) = X_{\mathrm{rms}}(r,t) \equiv \sqrt{\langle X''^2 \rangle(r,t)}.
\end{align}

\newpage

\bibliography{reference}
\bibliographystyle{aasjournal}
\end{document}